\documentclass[twocolumn,10pt]{revtex4}%
\usepackage[paperwidth=210mm,paperheight=297mm,centering,hmargin=2cm,vmargin=2.5cm]{geometry}
\usepackage{amsfonts}
\usepackage{amsmath}
\usepackage{amssymb}
\usepackage{bm}
\usepackage{graphicx}
\usepackage{color}
\usepackage{soul}
\usepackage{epsfig}%
\usepackage[dvipsnames]{xcolor} 
  \definecolor{bleu_cite}{RGB}{0,0,255}
\usepackage[colorlinks=true,allcolors = black,citecolor=bleu_cite,  ]{hyperref} 
\usepackage{natbib}

\usepackage{siunitx}

\begin{document}
\title{Quasi-condensation of bilayer excitons in a periodic potential}

\author{Camille Lagoin$^1$, Stephan Suffit$^{2}$, Kenneth West$^3$, Kirk  Baldwin$^3$, Loren Pfeiffer$^3$, Markus Holzmann$^4$ and Fran\c{c}ois Dubin$^{1}$} 
\affiliation{$^1$ Institut des Nanosciences de Paris, CNRS and Sorbonne Universit{\'e}, 4 pl. Jussieu,
75005 Paris, France}
\affiliation{$^2$ Laboratoire de Materiaux et Phenomenes Quantiques, Universite Paris Diderot, 75013 Paris}
\affiliation{$^3$ PRISM, Princeton Institute for the Science and Technology of Materials, Princeton University, Princeton, NJ 08540, USA}
\affiliation{$^4$ Univ. Grenoble Alpes, CNRS, LPMMC, 38000 Grenoble, France}

\begin{abstract}
We study two-dimensional excitons confined in a lattice potential, for high fillings of the lattice sites. We show that a quasi-condensate is possibly formed for small values of the lattice depth, but for larger ones  the critical phase-space density for quasi-condensation rapidly exceeds our experimental reach, due to an increase of the exciton effective mass. On the other hand, in the regime of a deep lattice potential where excitons are strongly localised at the lattice sites, we show that an array of phase-independent quasi-condensates, different from a Mott insulator, is realised.

\end{abstract}

\maketitle

The phase diagram of bosonic particles exploring a lattice potential is commonly described in terms of two
essential parameters, the inter-particle interaction strength $U$ giving the increase of energy when two particles occupy the same lattice site, and the particle tunnelling strength $J$, determining the hopping of a particle to a neigbouring site \cite{Greiner_PhD,Salomon_2010,Lewenstein_book}. Within the minimal Bose-Hubbard model, two antagonist regimes emerge at zero temperature. When tunnelling is dominant, a superfluid showing phase ordering over spatially
distinct sites can be reached, whereas above a critical ratio $U$/$J$ tunnelling is suppressed and bosons remain fixed to their lattice site. This results in an incompressible phase, a so-called Mott insulator, where phase coherence extending over various lattice sites is absent. Typically, varying the lattice depth or the lattice period, $U$ and $J$ are modified and the phase diagram is accurately explored \cite{Salomon_2010,Lewenstein_book}.
 
Seminal experiments conducted with ultra-cold atomic gases confined in optical lattices have provided model studies of the Bose-Hubard model, initially in three dimensions, and more recently, for effectively two and one dimensional systems \cite{Lewenstein_book}. In the solid-state, large efforts have also been dedicated to the physics of the two-dimensional Bose-Hubbard Hamiltonian, in particular for quantum simulation perspectives. To this aim, promising candidates include long-lived dipolar excitons of semiconductor bilayers. 
Such excitons are marked by a spatial separation imposed between Coulomb-bound electrons and holes. They are possibly engineered in coupled GaAs quantum wells  \cite{Combescot_ROPP} or in van der Waals assemblies of transition metal dichalcogenides monolayers \cite{Rivera_2018}. In the former case, excitons may be subject to an artificial lattice potential engineered by electrostatic gates \cite{Lagoin_2020,Remeika_2011} while in the latter case they naturally explore a periodic moire potential \cite{Seyler_2019,Tran_2019,Jin_2019}.  For both systems, the Bose-Hubbard model is expected to provide an accurate low-energy description.

To explore the collective phases accessible to dipolar excitons, GaAs double quantum wells provide 
an experimental toy model system, since excitons are then possibly manipulated in a regime of homogeneous broadening \cite{Dang_2020}, and spatially confined in on-demand potential landscapes \cite{High_2009,Grosso_2009,Winbow_2011,Kuznetsova_2015,Butov_review_2017,Shilo_2013}.
Using these degrees of freedom, we have recently mapped out the quasi-condensation crossover 
of bilayer excitons in box-like trapping potentials \cite{Anankine_2017,Dang_2019,Dang_2020}, determining the excitons equation of state and density fluctuations, and correlated these to the degree of spatial and temporal coherence at sub-Kelvin temperatures.

Here, we report experiments characterising  the quasi-condensation of GaAs bilayer excitons in a microscopic lattice potential. We show that a quasi-condensate is destroyed above a threshold lattice depth around a fraction of the chemical potential. We attribute the suppression of coherence to the renormalisation of the exciton  effective mass by the lattice potential which rapidly increases the critical phase space density of quasi-condensation. 
Above a threshold lattice depth, cold excitons enter then a normal phase, but without yet being spatially localised at the lattice sites. The localised regime is in fact obtained for lattice depths large compared to the chemical potential.
In this regime, we show that  an array of phase incoherent quasi-condensates localized at each site is possibly formed at large filling factors. We underline that this realisation differs from a Mott insulator. 

\begin{figure}[!ht]
  \includegraphics[width=.9\linewidth]{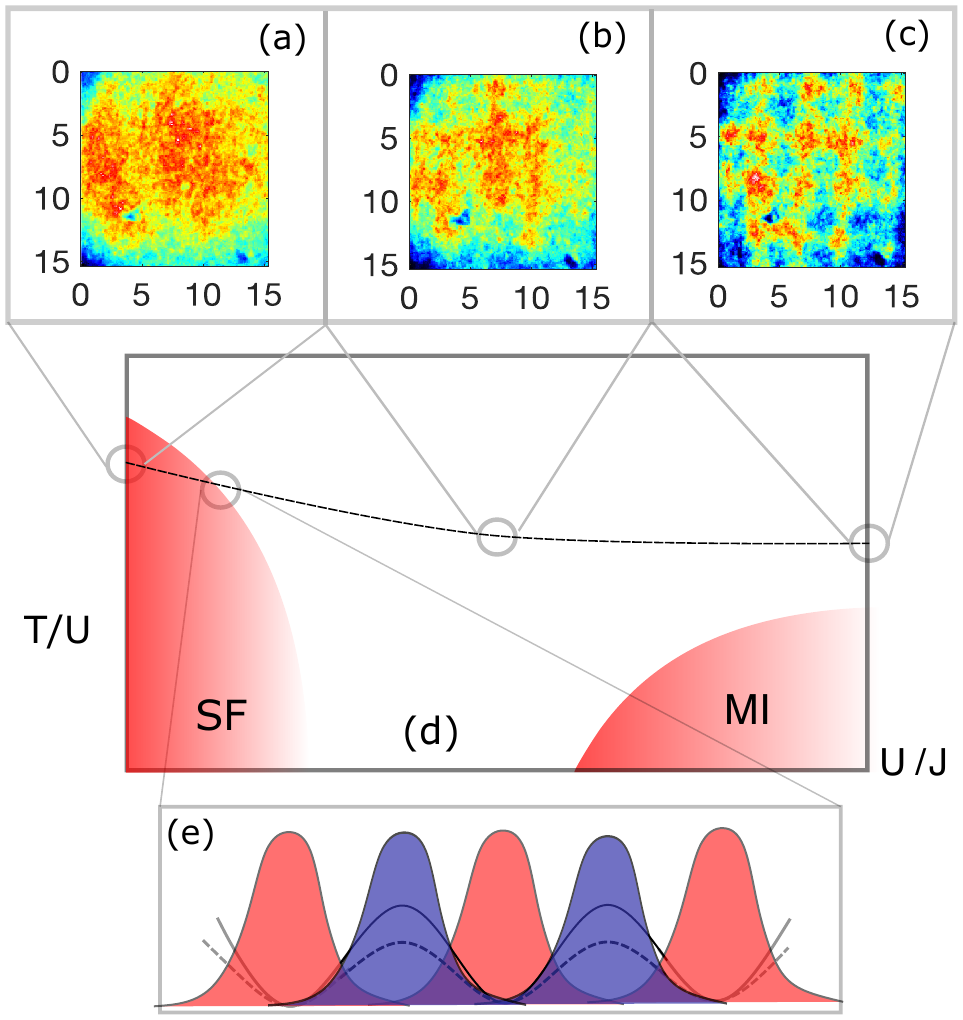} 
  \caption{(a-c) Real images of the photoluminescence for $V_0$=0 (a), 1.5 meV (b) and 3 meV (c). Scales are given in $\mu$m. (d) Schematic phase diagram for the two-dimensional Bose-Hubbard model, as a function of the on-site interaction $U$, the tunnelling strength $J$ and the temperature $T$. At non-zero temperature superfluid and Mott insulating phases, SF and MI respectively, are separated by a normal phase where physical properties are classical. Our experiments, carried out at a fixed average density, explore the parameter space along the dashed curved line. (e) At the crossover between the SF and normal phases non-condensed excitons  (blue) are mostly located around the lattice barrier. This effectively increases the  lattice depth (solid line) compared to the native one (dashed line). Quasi-condensed excitons are depicted in red.}
\end{figure}

\textcolor{black}{We study a square artifical lattice with spatial period L=3 $\mu$m, at a bath temperature $T_b$ set to 340 mK (see Ref. \cite{Lagoin_2020} for the caracterization of our device). Indirect excitons are injected in the lattice potential using a 100 ns long laser excitation, repeated at 1.5 MHz, resonant with the direct exciton absorption. In the following, we focus onto excitonic properties at a fixed average density $n\sim$ 2 10$^{10}$ cm$^{-2}$, obtained 150 ns after extinction of the laser excitation for an average power set to 1.5 $\mu$W. The phase-space density $D$ is then around 12 corresponding to a chemical potential $\mu$ of about 0.8 meV, i.e. well above the critical value $D_c\sim$ 8 for exciton quasi-condensation \cite{Dang_2020}. Thus, we explore how the lattice depth influences quantum statistical signatures, by continuously varying the height of the potential barrier $V_0$ separating the lattice sites \cite{Lagoin_2020}.}

Figure 1.d illustrates the phases potentially accessible in our experiments: for small $U/J$, realized in the limit of vanishing lattice depth, a quasi-condensate phase marked by quasi-long range order in the spatial and temporal coherence is expected \cite{Salomon_2010,Prokofeev_2018}. \textcolor{black}{On the other hand, for a deep lattice potential, i.e. for large $U$/$J$, one possibly enters the regime where a Mott-insulating phase (MI), defined by the same fixed number of excitons per lattice site, is energetically favourable. We then note that MI phases are protected since their minimum excitation, resulting from an exciton hoping between two lattice sites, has an energy cost equal to $U$=$(\hbar^2/4\pi m_X a_0^2)$$\tilde{g}$, $m_X$ being the excitons effective mass, $a_0$ describing the spatial extension of the excitons wave-function in lattice sites while $\tilde{g}$$\sim$4 is a dimensionless parameter quantifying dipolar interactions between excitons \cite{Dang_2020}. For our experimental conditions we deduce that $U\sim$$k_B T_b$. Then MI phases are not accessible due to thermal excitations. Moreover, let us note that for our device we only have access to high-filling factors, with around 100  excitons per site. }

We now turn to discuss our experimental results in these different regimes.
Let us start with the quasi-condensate regime 
at vanishing $U/J$, by setting $V_0$ to 0. In this situation, we studied the degree of spatial and temporal coherence of the photoluminescence radiated by our device using
 a Mach-Zehnder interferometer where the photoluminescence field $\psi$ is recombined with itself, after a variable time delay $\tau$, or a spatial displacement $\textbf{r}$ are introduced.  Measuring the interference contrast at the interferometer output, we directly deduced the amplitude of the first-order correlation function $|g^{(1)}(\textbf{r},\tau)|\sim|\langle \psi^*(\textbf{r}_i,t)\psi(\textbf{r}_i+\textbf{r},t+\tau)\rangle_{r_i,t}|$. Here $\langle ... \rangle_{r_i,t}$  denotes the average over $t$, i.e. over the number of  realisations accumulated to produce one interferogram, as well as the average over the positions $\textbf{r}_i$ where the fringe contrast is evaluated, typically of the order of 10 $\mu$m to compute the contrast from 3 fringes (Fig.2.a-b).

\begin{figure}[!ht]
  \includegraphics[width=.9\linewidth]{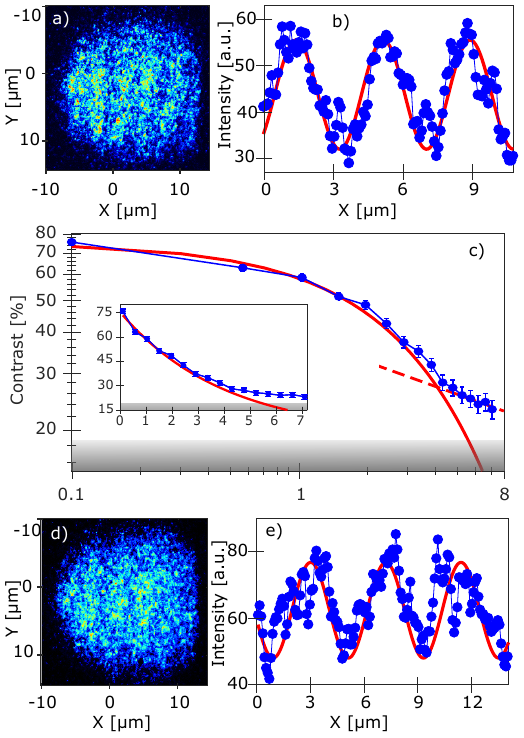} 
  \caption{(a) Interference pattern measured for $V_0$=0 and for (\textbf{r}=0, $\tau$=4.2 ps), together with the profile evaluated at the center of the image (b) leading to  $|g^{(1)}|$=28 $\%$. (c) Decay of the first-order time coherence $|g^{(1)}(0, \tau)|$ as a function of the time delay $\tau$. Experimental data are depicted by the solid blue points, while the red line shows an exponential decay with a characteristic time constant of 4 ps. The dashed line shows an algebraic time decay $\tau^{-\eta}$ with $\eta$=0.25. The inset presents the same measurements in linear scale \textcolor{black}{and the grey shaded area marks the limit of our experimental precision}.(d) Spatial interference of the photoluminescence measured for ($||$\textbf{r}$||$=2 $\mu$m, $\tau$=0), together with the interference profile evaluated at the center of the image (e). Thus we deduce an interference contrast $|g^{(1)}|$=23 $\%$. In (a-b) the interference period is set to 3 $\mu$m and 4.5 $\mu$m in (d-e).}
\end{figure}

Figure 2.c shows the variation of the photoluminescence first order time correlation function $|g^{(1)}(0,\tau)|$. We note that initially it decays exponentially with a time constant $\tau_c\sim$ 4 ps, before a slower decay is found for 4 $\lesssim\tau\lesssim$ 7 ps. Very recently we have reported such behaviour and shown that the initial exponential decay reveals the contribution of non-condensed excitons through the inelastic two-body collisional rate. Indeed, the exciton-photon coupling is linear so that the coherence of optically bright excitons is imprinted in the photoluminescence \cite{Monique_book}. Thus, we find here an excitons collisional rate, 1/$\tau_c$, in good agreement with our previous studies \cite{Dang_2020}. On the other hand, the slowly decaying  part for $\tau\gtrsim\tau_c$ marks the contribution of quasi-condensed excitons, which exhibit an algebraically decaying time coherence $\tau^{-\eta}$ \cite{Prokofeev_2018}. Importantly, here we find that $\eta\sim$0.25 which is compatible with the value predicted at criticality by the Berezinskii-Kosterlitz-Thouless theory \cite{Berenzinskii,KT}. Let us then note that our studies are realised for an exciton phase-space density $D\sim12$, very close to Ref. \cite{Dang_2020} where a very similar exponent was deduced.

We further studied the degree of spatial coherence by setting our interferometer such that $||\textbf{r}||\sim$2 $\mu$m and $\tau$=0. This spatial separation is over an order of magnitude beyond the classical limit set by the thermal de Broglie wavelength, and two times larger than our optical resolution. Figure 2.d reveals that we observe interference fringes signaling the buildup of quasi-long-range spatial coherence. In these experiments the interference contrast amounts to 23$\%$, compared to 75$\%$ at the spatial auto-correlation. From this difference we deduce that 1/3 of the optically bright excitons population contribute to the quasi-condensate \cite{Glauber_99}. \textcolor{black}{In previous works, we had shown that a quasi-condensate also includes optically dark excitons, that constitute around 2/3 of the total exciton population at 340 mK \cite{Beian_2017}. Thus, the fraction of quasi-condensed excitons amounts to 50(25) $\%$ : 58(42) $\%$ and 33 $\%$ of dark and bright excitons being  quasi-condensed, respectively. Let us note that the fraction of quasi-condensed dark excitons is extracted with a low accuracy since these are optically inactive so that their quantum statistical distribution is not measurable precisely. Nevertheless, the previous magnitudes well reproduce the ones we deduced in electrostatic traps for similar densities \cite{Anankine_2017}, in good agreement with path integral Monte-Carlo calculations \cite{Filinov}. Also, note that the amplitude of $|g^{(1)}|$ is of the same order for ($||\textbf{r}||\sim$2 $\mu$m, $\tau$=0) and (\textbf{r}=0, $\tau\sim$6 ps). This matching is expected since in both cases only the fraction of quasi-condensed bright excitons contributes to the interference signal.} 

We now turn to the opposite regime where a deep lattice potential is imprinted, $V_0$=3 meV$\gg\mu$. In this situation Fig.1.c shows that excitons are strongly localised in the lattice sites which have a characteristic spatial extension of about 1$\mu$m. This results in a modulation of the photoluminescence intensity, of around 20$\%$ along the horizontal axis of our device, with a 3 $\mu$m period \cite{Lagoin_2020}.  To study spatially extended time coherence in the photoluminescence, we studied $|g^{(1)}(0,\tau)|$ fixing the interference period to 4.5 $\mu$m, as in the measurements shown in  Fig. 2.d-e. Thus, the interference period is clearly distinguished from the one of the spatial modulation of the photoluminescence intensity due to the lattice potential.  Figure 3.a shows the variation of the interference contrast with $\tau$, which, unlike in Fig.2.c, is reduced to an exponential decay with a characteristic time of around 3 ps. 
We then note that for $V_0$=3 meV a slightly higher rate of inelastic collisions is deduced compared to $V_0$=0. We attribute this difference as the manifestation of an increased concentration of excess carriers \cite{Anankine_2018} when we impose a large difference between the potentials applied onto our gate electrodes, as necessary to engineer a deep lattice potential. As a result, excitons suffer additional collisions with free carriers, nevertheless, the photoluminescence is homogeneously broadened in this regime as well.

\begin{figure}[!ht]
  \includegraphics[width=.9\linewidth]{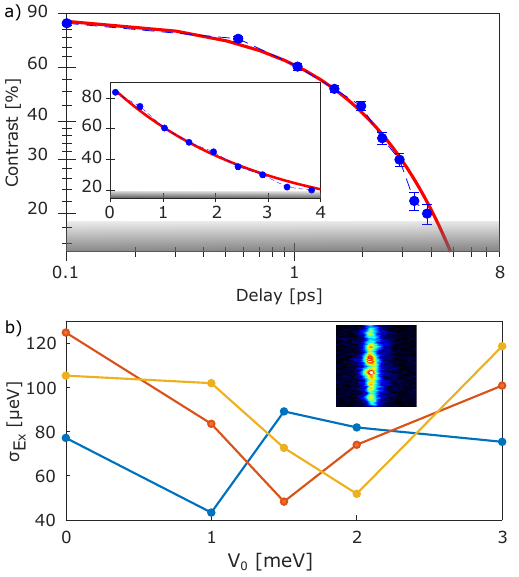} 
  \caption{(a) Decay of the first-order time coherence for $V_0$=3 meV in log-log scale. Experimental data are displayed by the blue points while the red line shows an exponential decay with a characteristic time equal to 2.8 ps. The inset provides the same results in linear scale \textcolor{black}{while the grey area underlines the limit set by the signal-to-noise ratio}. (b) Standard deviation of the photoluminescence energy measured at the position of 3 adjacent lattice sites as a function of the lattice depth $V_0$. Each measurement is evaluated from an ensemble of 10 realisations all performed under the same experimental conditions. The inset provides a spatially resolved spectrum where the emission from individual lattice sites is identified \cite{Lagoin_2020}.}
\end{figure}

Figure 3.a shows that in a deep lattice potential cold excitons do not exhibit any algebraically decaying time coherence, and therefore lack from quasi-long-range order \cite{Dang_2020,Prokofeev_2018}. This implies that the phase between excitons confined in each lattice site is randomly distributed. We actually expected such conclusion since the coupling between the lattice sites, controlled by the tunnelling coefficient $J$, decreases exponentially with the barrier height, like $e^{-(V_0/E_r)^{1/2}}$ where $E_r$=$\pi^2\hbar^2/(2mL^2)\sim$0.2 $\mu$eV \cite{Salomon_2010}. $J$ is then vanishingly small for $V_0\sim$3 meV. Moreover, the schematic phase diagram shown in Fig.1 recalls that for $V_0$=3 meV a Mott insulating phase is potentially accessible. As mentioned earlier, this is unlikely in our experiments carried out at a temperature comparable to the energy gap protecting Mott phases. To verify this expectation we studied the variation of density fluctuations in the lattice sites as a function of the lattice depth. Indeed a Mott insulator is signalled by the same  fixed number of particles in each site, so that density fluctuations are theoretically vanishingly small. For dipolar excitons density fluctuations are directly accessed by the energy of the photoluminescence which is governed by the strength of repulsive dipolar interactions between excitons \cite{Ivanov_2010,Schindler_2008,Rapaport_2009}. Fig.3.b compares the standard deviation of the photoluminescence energy $\sigma_{E_X}$ for 3 neighbouring lattice sites as a function of $V_0$. Overall, it shows that the photoluminescence energy is highly stable in our experiments, since it varies by only around 80 $\mu$eV. However, we do not observe any clear dependence of $\sigma_{E_X}$ over $V_0$, and thus no sign of a Mott insulating phase is detected in the deep lattice regime. \textcolor{black}{Moreover, we note that Fig.3.b indicates strongly that thermodynamic equilibrium is reached across the lattice potential since density fluctuations do not depend on $V_0$ \cite{Andreev14}.} 

\begin{figure}[!ht]
\includegraphics[width=.8\linewidth]{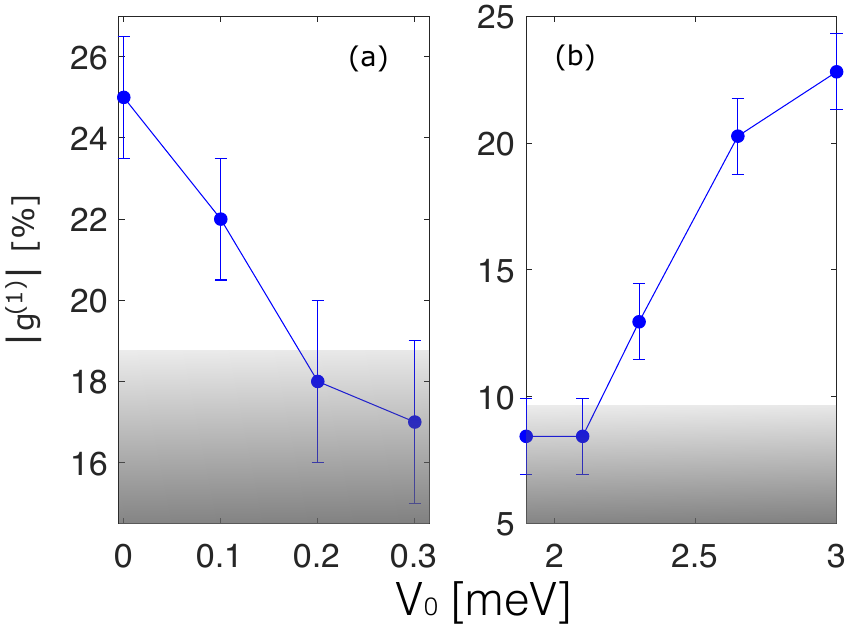} 
\caption{(a) First-order time coherence as a function of the barrier height $V_0$ in the weak lattice limit. The contrast measures the amplitude of the first-order time correlation function $|g^{(1)}(0,4.7ps)|$. (b) Local time coherence in the deep lattice limit. Here the contrast measures the amplitude of the intensity modulation along a bright interference fringe, i.e. along the weaker confining direction. The shaded areas display our instrumental resolution limited by the signal to noise ratio in these measurements. Experiments have all been performed at a bath temperature $T_b$=340 mK.}
\end{figure}

Figure 2 and 3 show that extended temporal coherence is destroyed when we pass from the regime where excitons are delocalised in the lattice potential ($V_0=0$) to the one where they are strongly localised ($V_0=3$ meV). To quantify this transition, we have measured the degree of temporal coherence as a function of the lattice depth, setting $\tau$ to 4.7 ps. Starting from a flat potential ($V_0$=0), Fig. 4.a shows that $|g^{(1)}|$ decays very rapidly when $V_0$ is increased by a few 100 $\mu$eV. Actually  $|g^{(1)}|$ is reduced to the amplitude it would reach by decaying exponentially at a rate $\tau_c$ when $V_0\sim$  0.2 meV. This magnitude signals that  time coherence is governed by inelastic two-body collisions only, manifesting that the fraction of quasi-condensed excitons has vanished. 

\textcolor{black}{The loss of coherence in a periodic potential is usually discussed in terms of arrays of condensates in single lattice sites connected by a Josephson coupling describing  quantum tunnelling \cite{Bradley84,Pitaevskii01,Burger02,Andreev17,Andreev14}. The latter decreases exponentially with the difference between the barrier height and the chemical potential \cite{Dalfovo96}. In our strongly interacting system, the barrier height seen by the quasi-condensate must also include the interaction with non-condensed excitons. Indeed, unlike in a flat confining potential, in a lattice non-condensed excitons are mostly localized around the barriers separating accessible sites \cite{HKN99} (see Fig.1.e). This periodic arrangement minimises the system energy and leads to an effective barrier height $V_0^*\approx V_0+2(\hbar/m)\tilde{g}(n_{nc}^{(max)}-n_{nc}^{(min)})$. Here, $n_{nc}^{(max/min)}$ are the maximum/minimum local density of non-condensed excitons. As previously discussed, their difference is of the order of half of the total population. Accordingly, we deduce that $V_0^*\sim\mu$ for $V_0\sim$ 0.2 meV $\ll\mu$, and a transition from a coherent condensate to incoherent array of micro-condensates can be expected forour parameter space  \cite{Bradley84,Pitaevskii01}. Alternatively, at finite temperatures the loss of coherence can be understood in terms of a renormalized effective mass $m^*_X$=$m_X/(1-V_0^{*2}/2\mu^2)$ for small lattice amplitudes ($V_0\ll\mu$) \cite{Moelmer}. For $V_0^*\sim\mu$ one has  $m^*_X\sim2 m_X$ so that the effective critical phase space density for quasi-condensation increases by a factor of two, from 8 to 16. It then exceeds our experimental value $D\sim$ 12, driving a transition from a coherent (quasi-uniform) condensate to an incoherent normal gas, as observed in Fig.4.a.}

\textcolor{black}{The experiments reported in Fig.4.b allow us to distinguish the two previous scenarii. There, we extracted the exciton time coherence locally, i.e. in the lattice sites, by ensuring that one bright interference fringe coincided with one row of lattice sites.} For each value of $V_0$ we then measured the intensity modulation along this particular bright fringe $C_i$, and compared it to the bare modulation of the photoluminescence in real space $C_m$ due to the localisation in the lattice. Thus, we directly deduce the average amplitude of $|g^{(1)}|$ for the photoluminescence radiated by the lattice sites only. It reads $|g^{(1)}|$=2$(C_i-C_m)/((1-C_i)(1+C_m))$, since the emission between the lattice sites does not yield any measurable interference signal for $\tau$=4.7 ps (Fig.3.a). \textcolor{black}{Figure 4.b reveals then that the interference contrast is vanishing for $V_0\lesssim$ 2 meV, the contrast increasing thereafter rapidly with $V_0$. For $V_0$ = 3 meV it reaches an amplitude  similar to the one for $V_0\sim$ 0. Accordingly, we deduce that for $V_0\lesssim$ 2 meV excitons lack extended temporal coherence, whereas coherence builds up for $V_0\gtrsim$ 2 meV providing a signature of microscopic quasi-condensates in the lattice sites for $V_0$ = 3 meV \cite{Dang_2020,Prokofeev_2018}. As shown in Fig.3, these coherent exciton droplets are independent from one another, i.e. with no defined phase relation. Figure 4 shows that the transition between this array of droplets and the extended quasi-condensate ($V_0\sim$ 0) passes then through a normal incoherent exciton gas which is not localised by the lattice potential (see Ref. \cite{Lagoin_2020}). Such transition does not correspond to the framework of Josephson coupled array of condensates, but rather suggests that coherence is destroyed due to a renormalisation of the exciton effective mass, which yields a critical phase-space density exceeding our experimental conditions.}

To conclude, we have experimentally studied the quasi-condensate crossover for bilayer excitons confined in a lattice potential. We have observed that a quasi-condensate is formed when the barrier height is vanishing, but rapidly destroyed when it is increased to around 0.2 meV. We have shown that this behaviour is consistent with a renormalisation of  the exciton effective mass by the lattice depth, which increases the critical phase-space density for the quasi-condensation crossover. On the other hand, in the deep lattice limit, we have observed that an array of phase incoherent quasi-condensates localized at the lattice sites develops. However, in order to reach the Mott insulator regime, we estimate that the period of the lattice has to be decreased to less than around 1 $\mu$m, 
\textcolor{black}{such that the effective trapping frequency is increased in the lattice sites. Thus, the strength of on-site interactions is enhanced \cite{Greiner_PhD} and becomes a few times larger than the thermal activation energy}. 

\section*{Acknowledgments}
Our work has been financially supported by the Labex Matisse, the Fondation NanoSciences (Grenoble), and by OBELIX from the French Agency for Research (ANR-15-CE30-0020). The work at Princeton University was funded by the Gordon and Betty Moore Foundation through the EPiQS initiative Grant GBMF4420, and by the National Science Foundation MRSEC Grant DMR 1420541.

\end{document}